\documentstyle[12pt,epsfig]{article}
\title{{\LARGE\bf Coherent e$^+$e$^-$ pair creation at high energy
muon colliders.}
  \thanks{Talk at the Workshop Studies on Colliders and
    Collider Physics at the Highest Energies: Muon Colliders at 10 TeV
    to 100 TeV, 27 September - 1 October, 1999 Montauk, New York, USA,
    be published by the American Institute of Physics.  }  }

\author{Valery Telnov, \\
  {\small\it Budker Institute of Nuclear Physics, 630090 Novosibirsk, 
  Russia}\thanks{email:telnov@inp.nsk.su}} 

\begin{document}
\newcommand{\EP}{\mbox{e$^+$}}
\newcommand{\EM}{\mbox{e$^-$}}
\newcommand{\EPEM}{\mbox{e$^+$e$^-$}}
\newcommand{\EMEM}{\mbox{e$^-$e$^-$}}
\newcommand{\GG}{\mbox{$\gamma\gamma$}}
\newcommand{\GE}{\mbox{$\gamma$e}}
\newcommand{\TEV}{\mbox{TeV}}
\newcommand{\GEV}{\mbox{GeV}}
\newcommand{\LGG}{\mbox{$L_{\gamma\gamma}$}}
\newcommand{\EV}{\mbox{eV}}
\newcommand{\CM}{\mbox{cm}}
\newcommand{\MM}{\mbox{mm}}
\newcommand{\NM}{\mbox{nm}}
\newcommand{\MKM}{\mbox{$\mu$m}}
\newcommand{\SEC}{\mbox{s}}
\newcommand{\CMS}{\mbox{cm$^{-2}$s$^{-1}$}}
\newcommand{\MRAD}{\mbox{mrad}}
\newcommand{\IND}{\hspace*{\parindent}}
\newcommand{\E}{\mbox{$\epsilon$}}
\newcommand{\EN}{\mbox{$\epsilon_n$}}
\newcommand{\EI}{\mbox{$\epsilon_i$}}
\newcommand{\ENI}{\mbox{$\epsilon_{ni}$}}
\newcommand{\ENX}{\mbox{$\epsilon_{nx}$}}
\newcommand{\ENY}{\mbox{$\epsilon_{ny}$}}
\newcommand{\EX}{\mbox{$\epsilon_x$}}
\newcommand{\EY}{\mbox{$\epsilon_y$}}
\newcommand{\BI}{\mbox{$\beta_i$}}
\newcommand{\BX}{\mbox{$\beta_x$}}
\newcommand{\BY}{\mbox{$\beta_y$}}
\newcommand{\SX}{\mbox{$\sigma_x$}}
\newcommand{\SY}{\mbox{$\sigma_y$}}
\newcommand{\SZ}{\mbox{$\sigma_z$}}
\newcommand{\SI}{\mbox{$\sigma_i$}}
\newcommand{\SIP}{\mbox{$\sigma_i^{\prime}$}}
\date{}
\maketitle
\begin{abstract}

  It is shown that at muon colliders with the energy in the region of
100 TeV the process of coherent pair creation by the muon in the field
of the opposing beam becomes important and imposes some limitations on
collider parameters.

\end{abstract}

One of the main advantages of muon colliders is that the muon is much
heavier than the electron and therefore the radiation (beamstrahlung)
in beam collisions is suppressed. The relative energy loss during
the beam collision $\Delta E/E  \propto EB^2/m^4$.

However, there is another process in beam collisions which may be
important for a high energy muon collider: it is the coherent \EPEM\ 
creation.  In this process the \EPEM\ pair is created by a virtual
photon in a strong field of the opposing muon beam $B \equiv |E|+|B|$.
The process of coherent pair creation is very important for \EPEM\ 
linear colliders~\cite{CHEN,TEL90}. This process has large probability
at
\begin{equation}
\kappa = (\omega/m_ec^2)(B/B_0) >1, \;\;\;\;  B_0=\alpha e/r_e^2 \sim
4.4\times 10^{13}\; \mbox{Gauss}.
\end{equation}
 At a 100 TeV muon collider the energy and
beam field are even higher than those at  linear \EPEM\
colliders. So, one can expect that this process will be 
important for muon collider as well, because naively the cross section of
this process depends only on $E, B, m_e$, but not on $m_{\mu}$.

  However, there is one effect in this process which makes a situation
at electrons and muon beams very different. In \EPEM\ collisions, the
maximum energy of virtual photons is almost equal to the electron
energy, while at $\mu\mu$ colliders the maximum photon energy depends
also on the mass of the produced system. This can be understood in the
following way~\cite{Ginz}.  The minimum value of the photon mass, which
corresponds to the case when the virtual photon has zero transverse
momentum~\cite{Budnev},
\begin{equation}
Q^2_{min}=-q^2_{min}=-(p-p^\prime)^2 \approx \frac{m^2\omega^2}{E(E-\omega)},
\end{equation}
where $m$ is the mass of beam particles. Also, the cross section of
\EPEM\ pair production is large only near the threshold $W^2 \sim
4m_e^2$. Besides, the cross section is negligible for $Q_{max}^2>
W^2$, i.e. $Q_{max}^2\sim m_e^2$. As a result, from the inequality
$Q_{min}<Q_{max}$ it follows
\begin{equation}
\omega < \gamma_{\mu} m_e c^2.
\end{equation}
So, only photons with the energy $\omega < \gamma_{\mu} m_ec^2 \sim
(1/200)E_{\mu}$ contribute to the process of coherent pair creation.

  Nevertheless, at the 100 TeV $\mu\mu$ collider even such ``low
energy'' photons can produce \EPEM\ pairs. Indeed, for $N=0.8\times
10^{12},\sigma_x=2\times10^{-5}$ cm, $\sigma_z=0.25$ cm, $E=50$ TeV
(``evolutionary'' $\mu\mu$(100) collider)

\begin{equation}
\kappa \sim \gamma_{\mu}\frac{B}{B_0} \sim 0.85.
\end{equation}
Here I took $B \sim eN/\sigma_x\sigma_z$, which is close to the maximum
effective beam field ($|B|+|E|$).

 The probability of \EPEM\ creation by the muon in the transverse
magnetic field per  unit length for $\kappa < 1$ ~\cite{Baier}
\begin{equation}
W \sim \frac{0.013\alpha^3 R^{5/2}}{r_e \gamma_{\mu}}
e^{-2\sqrt{3}/R},
\end{equation}
where $R = \gamma_{\mu}(B/B_0)$.\footnote{Here I distinguish $R$
  and $\kappa$ because $\kappa$ is approximately equal to
  $\gamma_{\mu}B/B_0$ while $R$ is equal to this expession by definishion.}
For $R=0.85$, $\sigma_z=0.25$ cm,
  $E_{\mu}=50$ TeV the probability of \EPEM\ pair creation by the muon
  during its life (about 1000 beam collisions)
\begin{equation}
p \sim 1000 W\sigma_z \sim 0.1.
\end{equation}
This is a large probability, the maximum that can be accepted. Further
two times increase of the  $R$  value will  lead to
a one order decrease of the luminosity.  Note, that in the process of the
\EPEM\ creation the muon loses about 1/200 of its energy that is much
larger than the energy spread at muon colliders ($\sim 10^{-4}$), so
the considered muon will no longer contribute  to the luminosity (due
to chromatic abberation).

Let us compare now coherent pair creation with beamstrahlung where the
muon can also emit of sufficiently hard photon.  We have seen that the
probability of  coherent pair creation is large when $\kappa \sim
(EBe\hbar)/(m_{\mu}m_e^2 c^5) > 1$. In  beamstrahlung, the muon is
``lost'' when the characteristic photon energy $E_{\gamma}/E \sim
(EBe\hbar)/(m_{\mu}^3 c^5) = \kappa (m_e/m_{\mu})^2 > \delta \sim
10^{-4}$ (see B.King's table  of the 100 TeV ``evolutionary'' muon
collider).  One can see that the expression for a beamstrahlung does
not contain $m_e$ and is smaller by a factor of $(m_{\mu}/m_e)^2 \sim
4\times 10^4$ than the characteristic parameter in  coherent pair
creation; however the upper limit on the beamstrahlung parameter is
also smaller by a numerically similar factor. This means than both
processes become important approximately at the same values of the
muon energy and beam field. Which process is more important depends on
the energy acceptance of the final focus system. For the considered
parameters the coherent pair creation is more important.
 
\vspace{0.5cm}
  The coherent \EPEM\ pair creation in  beam collisions is essential for
the 100 TeV muon collider and imposes some limitations on design parameters.  

\vspace*{0.5cm}

\end{document}